\documentstyle[prl,aps,multicol,epsf]{revtex}
\begin{document}
\draft
\widetext

\newcommand{\be}{\begin{equation}}
\newcommand{\ee}{\end{equation}}
\newcommand{\ber}{\begin{eqnarray}}
\newcommand{\eer}{\end{eqnarray}}

\title{Origin of synchronized traffic flow on highways and its dynamic
  phase transitions}

\author{H. Y. Lee$^{(1,2)}$, H.-W. Lee$^{(1)}$, and D. Kim$^{(1,2)}$}
\address{
$^{(1)}$Center for Theoretical Physics, Seoul National University \\
Seoul 151-742, Korea \\
$^{(2)}$Department of Physics, 
Seoul National University \\
Seoul 151-742, Korea}

\maketitle

\begin{abstract}
We study the traffic flow on a highway with ramps through
numerical simulations of a hydrodynamic traffic flow model. 
It is found that the presence of the external vehicle flux through ramps
generates a new state of recurring humps (RH).
This novel dynamic state is characterized by 
temporal oscillations of the vehicle density and velocity
which are localized near ramps, and found to be
the origin of the synchronized traffic flow
reported recently [PRL {\bf 79}, 4030 (1997)].
We also argue that the dynamic phase transitions between the free flow
and the RH state can be interpreted as a subcritical Hopf bifurcation.
\end{abstract}

\begin{multicols}{2}

\narrowtext


Experiences show that traffic flow has complicated properties.
The fact that automobile is one of main transportation
tools raises the traffic flow as one of the most important problems
for engineers \cite{Wohl}.
For physicists, on the other hand, traffic flow is
an interesting many-body problem of interacting vehicles.
Numerous experimental measurements revealed that 
traffic flow possesses qualitatively distinct dynamic states 
\cite{Treiterer}.
In particular, three distinct dynamic phases are observed
in highways \cite{Kerner96May}: 
the free traffic flow which is analogous to
laminar flow in fluid systems, the traffic jam state where
vehicles almost do not move, and the synchronized traffic flow 
which is characterized 
by complicated temporal variations of the vehicle density and velocity.

Paralleled with experiments, many physical models 
have been proposed \cite{Gerlough}.
Cellular automaton models \cite{Nagel} 
have been developed which simulate each individual vehicle, and
hydrodynamic models \cite{Kerner93,Helbing}
which provide macroscopic description of traffic flow.
Subsequent studies \cite{Schreckenberg,Kerner94} of the models
have explained many observed features of the
free flow and traffic jams in highways.
However, no satisfactory explanation for the 
synchronized flow is available to our knowledge.

Recently, Kerner and Rehborn
reported analysis of systematic measurements 
performed on German highways. As one of main results,
it was pointed out that the
synchronized flow is spatially localized
near ramps on highways \cite{Kerner97}. 
This observation motivated us to explore in this paper
effects of ramps 
on highway traffic flow.
Through numerical simulations of a hydrodynamic model,
we find that the presence of ramps generates
a new kind of traffic states
which becomes a spatially localized limit cycle 
of highway traffic flow under the constant external flux.
We examine properties of the novel state and show that
it is the origin of the synchronized flow.
 
In this work, we adopt the hydrodynamic model of highway traffic
flow proposed by Kerner and Konh\"{a}user \cite{Kerner93},
where the dynamic evolution is described by 
the Navier-Stokes-type equation of motion, 
\be
\rho \left({\partial v \over \partial t}+v{\partial v \over \partial x}\right)
={\rho \over \tau}(V(\rho)-v) -c^2_0 \ {\partial \rho \over \partial x}
+\mu {\partial^2 v \over \partial x^2} \ .
\label{eq:eqofmotion}
\ee
Here $\rho(x,t)$ is the local vehicle density, $v(x,t)$ the local
velocity, $V(\rho)$  the safe velocity that
is achieved in the time-independent and homogeneous traffic flow,
and $\tau,c_0, \mu$ are appropriate constants.
Eq.~(\ref{eq:eqofmotion}) is paired with
the modified equation of continuity \cite{Kerner95}, 
\be
{\partial \rho \over \partial t}+{\partial (\rho v) \over \partial x}
=q_{in}(t)\varphi (x-x_{in})-q_{out}(t)\varphi (x-x_{out}) \, ,
\label{eq:continuity}
\ee
where the source and the drain terms on the right hand side 
represent the external flux through an on-ramp and through an off-ramp, 
respectively \cite{singlerampcomment}.
Here $\varphi(x)$, describing the spatial distribution of the external flux,
is localized near $x=0$ and 
normalized so that $q_{in}(t)$ ($q_{out}(t)$) represents 
the total incoming (outgoing) flux.

To study effects of a single ramp, 
two ramps \cite{offrampcomment} are separated by a large distance 
$(|x_{in}-x_{out}|=L/2$ where $L$ is the system size)
and numerical simulations are performed 
with periodic boundary conditions.
The two-step Lax-Wendroff scheme is adopted as the main simulation
scheme, and its reliability is verified through comparison with
an alternative scheme:  
the classical fourth-order Runge-Kutta scheme applied to the time 
and the centered Euler scheme to the space \cite{Press}.
Simulations are carried out for many different sets of parameters and
qualitatively the same results are obtained.  
So for definiteness, we present results only for the following
choice of parameters : 
$\tau=$ 0.5 min, $\mu=$ 600 km/h, 
$c_0=54$ km/h, and
$V(\rho)=V_0  (1-\rho/\hat{\rho})/( 1+E (\rho/ \hat{\rho})^\theta )$ 
where the maximum density $\hat{\rho}=$ 140 vehicles/km, 
$V_0=120$ km/h, $E=100$, and $\theta=4$ \cite{Vcomment}. 
Concerning the discretization, spatial intervals of 
$\Delta x=37.8$ m and time intervals of
$\Delta t= 10^{-4}$ min 
are found to be suitable.
We choose the spatial distribution of the external flux as
$\varphi(x)=(2 \pi \sigma^2)^{-1/2} \exp{(-x^2/{2 \sigma^2})}$
with $\sigma=56.7$ m.

This model (\ref{eq:eqofmotion},\ref{eq:continuity}) was
investigated previously \cite{Kerner95} 
for the constant external flux $q_{in}(t)=q_{out}(t)=f$.
For small $f$, it was found that
an initially homogeneous flow $\rho(x,0)=\rho_h$,
$v(x,0)=V(\rho_h)$ evolves to slightly modified free flow
where homogeneous regions with different
densities are separated by narrow density-rising (or descending) regions
near the ramps, the so-called transition layers.
In contrast, for $f$ larger than a critical value,
a local avalanche-like
process occurs at the transition layer
and traffic jam appears spontaneously.  
This study, however, failed to
probe the synchronized traffic flow.

\begin{figure}
\epsfxsize=8cm \epsfysize=6cm \epsfbox{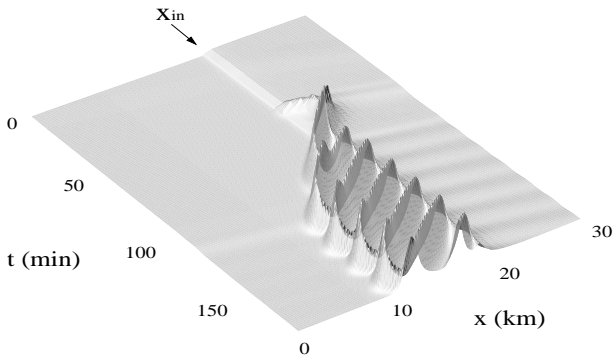}
\vspace{-1cm}
\caption{Birth and evolution of the RH state from the free flow is shown
  in this plot of the density profile ($\rho-\rho_{upstream}$)
  near the on-ramp ($x_{in}$=18.9 km) 
  for the average density of the system 
  $\rho_h$=22.4 vehicles/km $(\approx (\rho_{upstream}+\rho_{downstream})/2)$
  and the constant external flux $f=318$ vehicles/h  
  (the system size $L=75.6$ km).  
  Transition from the free flow to the RH state is triggered
  by a pulse-type perturbation applied at $t=50$ min
  with $\delta q_{in}$=318 vehicles/h, $\delta t=5$ min.
}
\label{fig:density3dplot}
\end{figure}

To find a clue to the missing third phase in traffic flow,
we pay attention to the experimental observation \cite{Kerner97} that
for a range of $f$, traffic flow can be either in the synchronized flow 
or in the free flow. This bi-stability suggests that 
the transition from one locally stable state to the other 
may require some triggering events.
So in our simulations, we apply a pulse-type perturbation
with a {\it finite} amplitude.
Specifically we first prepare a transition layer
by applying the constant external flux $f$ which is below
the critical point $f_c$ (for $f>f_c$, the stable free flow does not exist). 
Then a pulse of additional flux $\delta q_{in}$ is applied at
the on-ramp for a short duration $\delta t$. As a result,
a localized oscillating state appears from the
free flow \cite{initialbehaviorcomment} (Fig.~\ref{fig:density3dplot}).

After a transient period,
the localized oscillation becomes periodic in time. 
We observe that the Fourier spectrum of the oscillation shows sharp peaks at
each integer multiple of the basic frequency
$1/T$ ($\approx 0.068$ min$^{-1}$ in case of Fig.~\ref{fig:density3dplot}). 
In the simulation, it turns out that 
properties of this periodic asymptotic state,
such as the period and the oscillation amplitude,
are essentially independent of $\delta q_{in}$ and $\delta t$
as long as they are large enough to trigger the transition.  
This strongly suggests that the periodic oscillation 
is not a transient process with long decay time
but a limit cycle of Eqs.~(\ref{eq:eqofmotion},\ref{eq:continuity}).
For definiteness, we call this state of traffic flow
the {\it recurring hump} (RH) state in this paper,
which will be later compared with the synchronized flow. 

Limit cycles are generated in numerous examples 
of nonlinear autonomous systems 
(that is, systems without explicitly 
{\it given} time dependence) \cite{Jackson,Orszag}.
In the present example of traffic flow, 
the limit cycle can be characterized as 
a {\it self-excited} ({\it autocatalytic}) oscillator
(see Sec. 5.6 in Ref.~\cite{Jackson}),
where {\it constant} external flux serves as
a source of periodically generated excitations (humps).
Excitations are, however, relaxed within a localized region.
When the upstream vehicle
density is lower than the critical value $\rho_{cr} (\approx 25$ vehicles/km
for our parameter choice), a localized inhomogeneity decays away
in the homogeneous traffic environment unless the amplitude of the 
inhomogeneity is larger than a critical magnitude \cite{Kerner94}.
Thus humps cannot survive far away from the on-ramp if its size
does not exceed the critical magnitude.
In this way, the localization can be achieved.

\begin{figure}
\epsfxsize=8cm \epsfysize=3.7cm \epsfbox{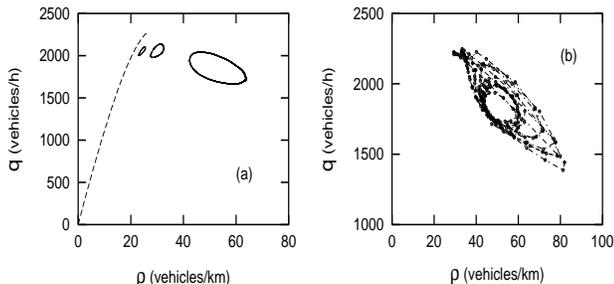}
\caption{(a) Density-flow diagrams for the average density
  $\rho_h=22.4$ vehicles/km, $f=318$ vehicles/h
  measured at $x-x_{in}=0$ km (the largest loop), $0.9$ km,
  and $3.8$ km (the smallest loop), respectively.
  The dashed line represents
  the free flow diagram.
  (b) Density-flow relation at the on-ramp under the presence of
  random fluctuations in the external flux. Data are shown for
  $q_{in,out}(t)=318$ vehicles/h $+\delta q_{in,out}(t)$ where
  $\delta q_{in,out}(t)$ is a random constant function
  in each time interval of $T_p=5$ min. At the end of each interval,
  the values of two independent random fluctuations
  $\delta q_{in}(t), \delta q_{out}(t)$ are
  reset by new random numbers uniformly distributed in the range
  $[-95,95]$ vehicles/h.
}
\label{fig:densityflow}
\end{figure}

The character of the localized oscillation becomes evident 
in the density-flow diagrams ($\rho(x,t)$ vs.$q(x,t)\equiv$ 
$\rho(x,t)$$v(x,t)$).
In contrast to a straight line for the free flow, the density-flow
relation for the RH state forms a closed loop at $x=x_{in}$
(Fig.~\ref{fig:densityflow}(a)),
which implies the periodicity of the oscillation and also 
the phase difference in oscillation between $\rho(x,t)$ and $q(x,t)$.
As $x$ moves downstream, 
the loop deforms gradually to a smaller loop
and eventually joins the free flow 
(Fig.~\ref{fig:densityflow}(a)),
which is a consequence of the localization.
We also examine the effect of randomly fluctuating 
external flux. 
Figure~\ref{fig:densityflow}(b) shows that though
the exact periodicity is lost, the oscillation itself
is still stable under random fluctuations. 

\begin{figure}
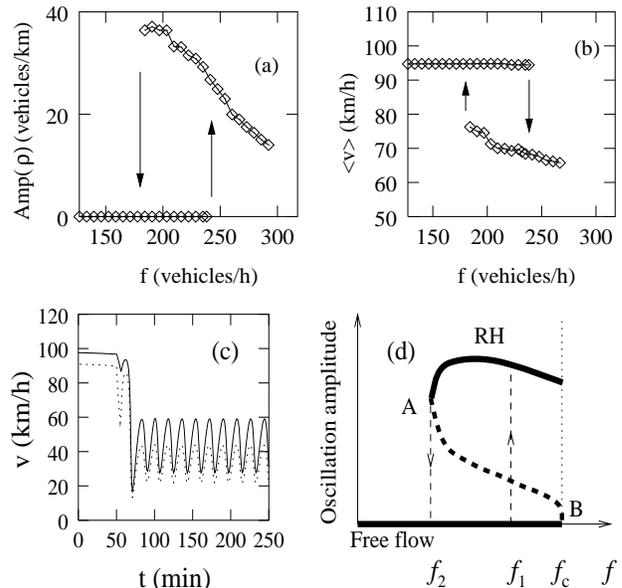

\epsfxsize=8cm \epsfysize=3.7cm \epsfbox{f3.epsi}
\vspace{0.3cm}
\epsfxsize=8cm \epsfysize=3.7cm \epsfbox{fsyn.epsi}
\vspace{0.1cm}
\caption{(a) Amplitude 
  of density oscillation measured at the on-ramp
  for $\rho_h=22.4$ vehicles/km.
  The zero amplitude implies free flow. Discontinuous jumps
  occur at $f_1 \approx 241$ vehicles/h and 
  $f_2 \approx 184$ vehicles/h.
  (b) $\langle v \rangle$ (Eq.~(\protect\ref{eq:averagevelocity})) 
  as a function of $f$ with $R=7.6$ km.
  Though the precise value of $\langle v \rangle$ depends on $R$, 
  the presence of discontinuities is universal.
  (c) Synchronized oscillations in a two-lane system. 
  The figure shows the temporal variations of the velocities at 
  the on-ramp for the lane 1 (solid line) and for the lane 2 (dotted line).
  A pulse of the external flux introduced at $t=50$ min induced 
  the transition to the RH state. Notice that the variations are 
  synchronized in both lanes.
  (d) Schematic diagram of the bifurcation scenario. 
  The point A corresponds to a turning point and B to 
  a subcritical Hopf bifurcation point.
  While the transition at $f_2$ is spontaneous,
  the transition at $f_1$ should be aided by an external
  triggering event. As a result, the value of $f_1$
  depends on the strength of triggering events. For $f>f_c$, 
  the free flow loses its stability spontaneously. 
}
\label{fig:transition}
\end{figure}

 Below we investigate
the transition from the free flow to the RH state.
For definiteness, we fix the perturbation, 
$\delta q_{in}=159$ vehicles/h, $\delta t=6$ min
and apply it to the transition layer generated by $f$ 
(Fig.~\ref{fig:transition}(a)). 
For small $f$, the free flow survives 
the perturbation. For $f$ larger than a critical value $f_1$,
however, the {\it finite}-amplitude RH state is induced.
We emphasize that a {\it finite} perturbation is essential
for the transition.
As the perturbation becomes weaker, $f_1$ becomes
larger and for $\delta q_{in}\delta t \rightarrow 0$, the transition
to the RH state does not occur for the whole range of
$f$ smaller than $f_c$ \cite{Kerner95}.
For the backward transition, on the other hand,
it turns out that
it may occur even without finite amplitude
perturbations.
As $f$ decreases adiabatically from $f>f_1$,
the amplitude of the RH state varies as in Figure~\ref{fig:transition}(a).
The system first follows its old path. 
Below $f=f_1$, however, the system still remains in the RH state
instead of going back to the free flow. 
The RH state is maintained until
$f$ reaches a lower critical value $f_2$,
where the transition to the free flow occurs \cite{lowercriticalpointcomment}.
We mention that transitions between the RH state and the 
free flow show the same hysteresis as measured 
in highways \cite{Kerner97}.
The influence of the transition on traffic flow becomes clear
in the following natural order parameter:
the spatio-temporal average velocity, 
\be
\langle v \rangle \equiv { 1 \over T R}
  \int_{t}^{t+T} dt' \int_{-R/2}^{R/2} dx \ v(x+x_{in},t') \ ,
\label{eq:averagevelocity}
\ee
where $T$ is the period of the RH state and $R$ is
the size of the averaging range. 
In Figure~\ref{fig:transition}(b),
$\langle v \rangle$ makes discontinuous jumps
at the transition points $f_1$ and $f_2$ \cite{nojumpinfluxcomment}.

Another interesting property of the RH state appears in multi-lane
situations. To demonstrate this, we extended
the traffic equations to a two-lane system.  
The equation of motion~(\ref{eq:eqofmotion}) and 
the continuity equation~(\ref{eq:continuity}) apply 
to each lane $i=1,2$. We assume that ramps are connected to the lane 2
and so the source and drain terms appear in the continuity
equation for the lane 2 only. 
We simulate the inter-lane interaction effect in a minimal way
by introducing to the continuity equations
lane-change terms $(\partial \rho_i / \partial t)_{ch}$
that account for the inter-lane flux due to the lane change of
vehicles.
For the simple choice,  
$(\partial \rho_i / \partial t)_{ch}= a(\rho_j(x,t)-\rho_i(x,t))  \ 
 (i\neq j)$, 
it is found that when the flow in the lane 2 makes the transition
to the RH state, it is accompanied by appearance of 
the synchronized oscillations of the velocity and the density
in the lane 1 (Fig.~\ref{fig:transition}(c)).
This property of the synchronization is examined for different
functional forms of $(\partial \rho_i / \partial t)_{ch}$ as well
since its precise form is not well determined yet.
Qualitatively same results are recovered in all cases, 
which demonstrates that the synchronization is a generic property
of the RH state in multi-lane situations \cite{hylee}.
It is worth commenting that 
this kind of synchronization phenomena is 
a common property in many examples of self-exciting systems 
(see Sec. 5.13 in Ref.~\cite{Jackson}).

Now it should be noticed that many properties of the RH state
are identical
to those of the synchronized flow \cite{Kerner97},
e.g., the discontinuous transition from the free flow to the synchronized flow
induced by localized perturbations of finite amplitudes,
hysteresis, stability of synchronized flow (hours of self-maintenance),
gradual spatial transitions from synchronized flow to free flow, 
and synchronized oscillations.
Therefore, we conclude that
the RH state is the origin of the synchronized traffic flow.

 We interpret our results within the standard framework
of nonlinear dynamics. The free flow corresponds to a point attractor.
On the other hand, many features of the RH state, such as
stability and discontinuous transitions, assert that
the RH state corresponds to a stable limit cycle. 
Hysteresis implies bi-stability for certain range of $f$.
The spontaneous backward transition from the limit cycle
to the point attractor means that the lower end of  
the bi-stable region is $f_2$. On the other hand,
the necessity of finite perturbations for the forward transition
suggests that the upper end of the bi-stable region
goes above $f_1$ and in fact extends to $f_c$.
Discontinuous transitions in both directions indicate
that the limit cycle is not connected to the point attractor
(Fig.~\ref{fig:transition}(c)).
The discontinuity also suggests that there should exist
still another asymptotic state, which 
serves as a boundary  between the basins of the attraction 
toward the two locally stable asymptotic states.
One plausible candidate for the boundary is 
an unstable limit cycle that is connected to 
the stable limit cycle at $f=f_2$ and
also to the fixed point at $f=f_c$ 
(Fig.~\ref{fig:transition}(c)).
In this case, the whole transition behaviors are results of
a {\it turning point} ($f_2$) combined with
a {\it subcritical Hopf bifurcation} ($f_c$).

Lastly we briefly discuss the synchronized flow far away from ramps
reported in Refs.\cite{Kerner96May,Kerner97}. 
There is one important difference between the synchronized flow 
near ramps and that far away from ramps.
While the former is a non-decaying state of oscillations, 
the latter appears only as a transient process; after some upstream movement
since its creation, it 
either disappears or transforms into a jam \cite{Kerner97}.
Therefore, the synchronized flow far away from ramps is
not a stable dynamic phase of traffic flow, 
and is not studied in this paper. The appearance of this
transient process requires further investigations. 
  
In summary, we find that there exists recurring hump (RH)
state in highway traffic flow with ramps. 
In this state, the density and the flow
oscillate periodically and the oscillations are
localized near the on-ramp.
The RH state is a stable limit cycle of the nonlinear traffic
equations~(\ref{eq:eqofmotion},\ref{eq:continuity}).
The transition between the free flow and the RH state
is discontinuous and shows hysteresis. 
Many features of the RH state are identical to those of
the synchronized flow and thus we conclude that the RH state
is the origin of the synchronized flow observed
in real highways \cite{Kerner96May,Kerner97}.   

\vskip 0.2cm
We acknowledge helpful discussions with Jysoo Lee, K. Chon, 
and M. Y. Choi.
HYL thanks the Daewoo Foundation
for financial support.
This work is supported by the Korea Science and Engineering
Foundation through the SRC program at SNU-CTP,
and by the Ministry of Education through
BSRI-97-2420.

\end{multicols}


\begin{references}
\bibitem{singlerampcomment} When an on-ramp and an off-ramp are
  closely positioned, and the influx through the on-ramp 
  is much larger than the outflux through the off-ramp, 
  this ramp pair can be treated as a single on-ramp.
\bibitem{offrampcomment} With periodic boundary condition, the presence of 
  the balanced off-ramp ($q_{out}=q_{in}$) is essential.
  Without it, the total vehicle number in the system
  varies with time, which prohibits investigation of asymptotic states.
\bibitem{Vcomment} The form of $V(\rho)$ is
  chosen so as to be consistent with available free flow data \cite{Kerner97}
and also with large density asymptotic behaviors. 
But its exact form is not important.
\bibitem{initialbehaviorcomment} In Figure~\ref{fig:density3dplot}(a),
  the initial direction of the first hump
  movement varies depending on the parameter choice and the strength
  of the perturbation. For sufficiently
  large perturbations, however, the initial direction is toward
  upstream.
\bibitem{lowercriticalpointcomment} For $f$ very close to $f_2$,
  oscillation patterns become complex 
  with the exact periodicity lost. Currently, we do not
  know the precise origin of the complexity.
\bibitem{nojumpinfluxcomment} Such discontinuities cannot be detected
  in the spatio-temporal average flow $\langle q \rangle$, because 
  the RH is connected to homogeneous regions, whose densities
  are essentially not altered by the appearance of the RH state
  if the system is sufficiently large, and hence the equation
  of continuity (\protect\ref{eq:continuity}) forbids
  finite change in the average flow upon the infinitesimal
  change in $f$.
\end{references}
\end{document}